\documentclass[a4paper,11pt]{article}
\usepackage{pos}
\usepackage{float}
\usepackage{xcolor}  % for \textcolor
\usepackage{soul}   % for strikethrough \st{}
\usepackage{csquotes}          %enquote
\usepackage{tikz}        % for words on top of slides

\newcommand{\overlayText}[5]{
  \begin{tikzpicture}
    \centering
    \node at (0, 0) {\includegraphics[width=#1,keepaspectratio,origin=c]{#2}};
    \node at (#3,#4) {#5};
  \end{tikzpicture}
}

\newcommand{\eqnr}[1]{Eq.~\ref{#1}}
\newcommand{\Fig}[1]{Fig.~\ref{#1}}

\title{Hybrid Charmonium at Finite Temperature}
\author*[a]{Juan Andr\'es Urrea-Ni\~no}
\author[a]{Ryan Bignell}
\author[a]{Ruaidhrí Campion}
\author[a]{Sin\'ead M. Ryan}

\affiliation[a]{School of Mathematics, Trinity College Dublin, Dublin 2, Dublin, Ireland\\and\\Hamilton Mathematics Institute, Trinity College Dublin, Dublin 2, Ireland }

\emailAdd{urreanij@tcd.ie}
\emailAdd{bignellr@tcd.ie}
\emailAdd{campioru@tcd.ie}
\emailAdd{ryansin@tcd.ie}

\abstract{Drawing upon well established zero-temperature techniques, we present, for the first time in a lattice calculation, insight into the fate of the $1^{-+}$ exotic charmonium state at finite temperature. Specifically, using anisotropic \textsc{Fastsum} ensembles we employ distillation with a wide operator basis which has been extensively used at zero-temperature by the Hadron Spectrum Collaboration to study the charmonium spectrum. The constant contribution to some finite-temperature temporal correlation functions requires particular care with the extended operator basis common to distillation setups and we discuss this effect. As an alternative to derivative based extended operators, we also consider the use of optimal distillation profiles at finite temperature for the first time. Finally, we remark on the temperature dependence of the 
$1^{-+}$ spectral function by consideration of the reconstructed correlator method.}

\FullConference{
}

\begin{document}
\maketitle

\section{Introduction}
States with quantum numbers not allowed in the quark-constituent model are referred to as exotic and have been the focus of many theoretical and experimental studies. The so-called XYZ-states, some of which are in the charmonium energy region, were detected and confirmed by Belle, BaBar and BESIII but their composition and dynamics are not yet fully understood~\cite{Brambilla-2020, Godfrey-2008}. These and other exotic states can be directly studied in lattice QCD, which is in principle systematically improvable and also allows for the exploration of mass and temperature dependence. In particular, the exotic charmonium state allowed by QCD with $J^{PC} = 1^{-+}$ has been extensively studied in lattice QCD at zero temperature by different collaborations~\cite{Liu-2012, Shi-2024} to establish its mass and decay dynamics. In Ref~\cite{Liu-2012} a convincing case is made for the identification of this state as a hybrid meson based on an analysis of operator overlaps. The study of these overlaps highlighted that in the large operator basis used, those subduced from continuum spin-1 operators with non-trivial gluonic content overlapped significantly on to the ground state of the $T_1^{-+}$ channel. 

In this work we turn our attention to the finite-temperature setting and investigate how zero-temperature spectroscopy techniques can be used (with some caveats) to study the fate of the hybrid meson as temperature increases. The pattern of melting and suppression of states in medium may offer an additional insight to the nature and structure of strong matter and the first steps in this direction for hybrid states is explored here.
\section{Methods and ensembles}
In lattice spectroscopy, states are classified according to the symmetries of the cubic group $O_h$ and its irreducible representations (irreps) $A_1, T_1, A_2, T_2, E$. By subduction and using the operator overlaps, as discussed above, these states can be related to states with continuum spin. Our goal is to use well-established zero-temperature spectroscopy techniques at finite temperatures to better understand the $1^{-+}$ charmonium state. A particularly robust approach is presented in \cite{Liu-2012}, where a wide basis of derivative-based meson operators~\cite{Dudek-2008} is used together with distillation~\cite{Peardon-2009}. The operators with lattice angular momentum quantum numbers, $T_1^{-+}$, for the state of interest here, are built from operators with fixed $J^{PC}$ in the continuum, in this case the relevant ones being $1^{-+}$, $3^{-+}$ and $4^{-+}$. 
While we expect that the ground state $T_1^{-+}$ can be identified as a continuum $1^{-+}$, nonetheless it is a good idea to use as many sensible operators as possible so that a GEVP calculation~\cite{Luscher-1990, Blossier-2009} can properly resolve all the relevant radial excitations in the one-particle case. Following this approach, we have 18 different operators using up to three spatial derivatives for the $T_1^{-+}$ hybrid charmonium. While these operators have the correct lattice angular momentum as well as parity and charge conjugation quantum numbers, there are two additional symmetry constraints to be considered.
\par
First, we have to work with periodic boundary conditions in time for the gauge fields. Given a two-point temporal correlation function between two meson operators $C(\tau) = \left \langle \mathcal{O}_i(\tau) \mathcal{O}_j(0)  \right \rangle$ over a total temporal extent $N_\tau$, then $C(N_\tau - \tau) = C(\tau)$ if both operators $\mathcal{O}_i$ and $\mathcal{O}_j$ have the same behaviour under time reversal or $C(N_\tau - \tau) = -C(\tau)$ if they have opposite behaviour. As explained in \cite{Bailas-2018}, this makes the extraction of energies from the GEVP eigenvalues more complicated if we consider a correlation matrix containing operators with different time-reversal symmetries. For this reason we take all operators for a given lattice irrep $R^{PC}$, $R=A_1,\, A_2,\, E,\, T_1,\, T_2$, and further classify them according to their time reversal symmetry. While this reduces the size of the correlation matrices available it allows us to extract energies from the periodic eigenvalues of the GEVP in the usual way. 
\par
Second, the use of finite temperature can introduce a constant contribution to two-point temporal correlation functions of any quantum numbers based on the meson operator used. In particular, for a meson operator of the form $\bar{q}\, \Gamma\, q$, the corresponding correlation function will have a constant contribution if $\{ \gamma_4, \Gamma \} \neq 0$~\cite{Umeda-2007, Ohno-2011, Arikawa-2025}. We could attempt to get rid of this constant contribution by applying a derivative in time or defining a midpoint-subtracted correlation $\tilde{C}(\tau) = C(\tau) - C\left( \frac{N_\tau}{2} \right)$ as proposed in \cite{Umeda-2007}. However, for small enough values of $N_{\tau}$ the correlation has not decreased enough and there are large numerical cancellations when doing the midpoint subtraction. To avoid dealing with numerical precision problems, we choose to further classify the operators into two groups: those which anti-commute with $\gamma_4$ and those which do not. This last sub-classification of operators yields a list of operators for each choice of $R^{PC}$, time-reversal and $\gamma_4$-anti-commutation. By starting with a large list of operators for all $R^{PC}$, the reduction in basis size is not too dramatic for most irreps and still allows for a variational approach to spectroscopy calculations.
\subsection{Ensembles}
This study uses the \textsc{Fastsum} collaboration Generation 2 ensemble configurations~\cite{Aarts:2014nba,Aarts:2020vyb,aarts_2023_8403827} which have a Symanzik-improved anisotropic gauge action with tree-level tadpole-improved coefficients with $N_f = 2 + 1$ flavours of anisotropic clover-improved Wilson fermions with stout-smeared links~\cite{Peardon-2004}. The tuned anisotropy is $\xi = \frac{a_s}{a_\tau}\approx 3.5$, with $a_s \approx 0.12$ fm and $a_\tau^{-1} \approx 5.7$ GeV. The spatial volume is $24^3$ for all temperatures, with temporal extents $N_\tau = 16, 20,24, 28, 32, 36, 40, 128$. In Table \ref{table:Gen2} we show the different temperatures in MeV and their ratio to the (pseudo)critical temperature $T_c$ set by the renormalised chiral condensate~\cite{Aarts:2020vyb}. The zero-temperature ensemble ($N_\tau = 128$) is the Hadron Spectrum (HadSpec) collaboration ensemble used for their charmonium calculations in \cite{Liu-2012}, where the pion mass is $\approx 385$ MeV. 

\begin{table}[H]
\begin{center}
\begin{tabular}{r|cccccccc}
$N_\tau$     & $128$ & $40$  & $36$  & $32$  & $28$  & $24$  & $20$  & $16$  \\ \hline
$T$ (MeV) & $44$  & $141$ & $156$ & $176$ & $201$ & $235$ & $281$ & $352$ \\
$T/T_c$   &   $0.243$    &   $0.777$    &   $0.864$    &  $0.972$     &   $1.11$    &   $1.30$    &  $1.56$     &  $1.94$    
\end{tabular}
\end{center}
\caption{Different temporal extents and corresponding temperatures in the Generation 2 ensemble.}
\label{table:Gen2}
\end{table}

\section{Spectrum calculation and reconstructed correlators}
\label{sec:Spectrum}
\begin{figure}
    \centering
    \includegraphics[width=0.8\linewidth]{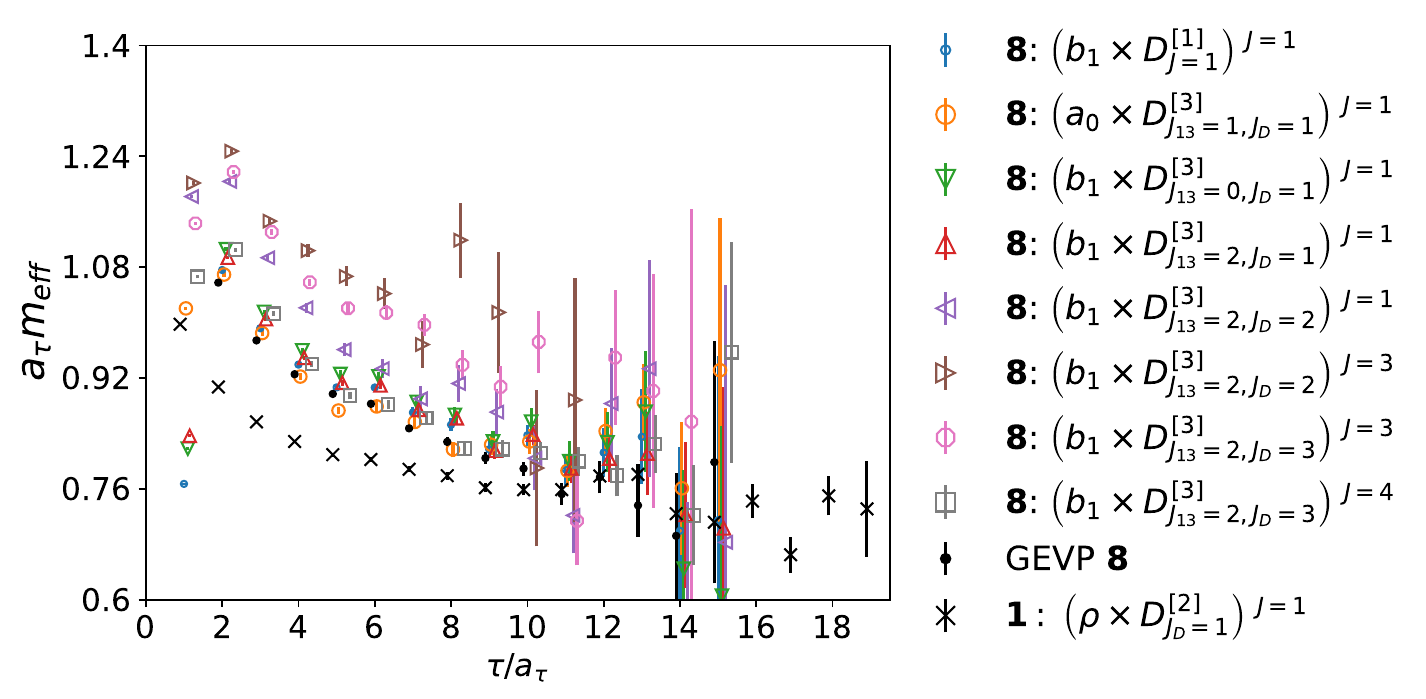}
    \caption{Effective masses of the ground state $T_1^{-+}$ $\left( 1^{-+} \right)$ charmonium at $N_\tau = 40$ using different operators labeled by the convention of \cite{Dudek-2010}. The \textbf{8} and \textbf{1} mean the operator is either in the group of eight operators with $\{ \Gamma, \gamma_4 \} \neq 0$ or the single operator with $\{ \Gamma, \gamma_4 \} = 0$ as explained in Sec. \ref{sec:Spectrum}.%\tcr{Ryan: Can we just not plot the noisy points? Additionally remove the title, make axis labels \& markers bigger and make the marker sizes a little bigger (much bigger in the legend). Instead of using stars for all the 8 operators, we should use hollow but different symbols.}
    } 
    \label{fig:T1mpComparison}
\end{figure}

The focus in this study is the $T_1^{-+}$ channel, for which we originally had the 18 operators also used in \cite{Liu-2012}. Enforcing same time-reversal symmetry results in two groups of operators, each with nine operators. We found that the individual operators of one group yield effective masses almost identical to their corresponding partner in the opposite time-reversal channel. This is not unexpected, as for the $A_1^{-+}$ the operators $\gamma_5$ and $\gamma_4 \gamma_5$ have opposite time-reversal symmetry and are often found to produce almost identical effective masses in the literature when using connected-only correlations, e.g.~\cite{Urrea-2025}. We choose the time-reversal even operators and further divide these into two groups: eight operators with $\{ \Gamma, \gamma_4  \} \neq 0$ and one with $\{ \Gamma, \gamma_4  \} = 0$. At first glance this seems problematic since only one operator has all the desired symmetries and therefore a GEVP approach is not possible. One might think it is worth trying a GEVP with the eight operators of the other group and deal with the numerical cancellations. However, this turns out not to be necessary. \Fig{fig:T1mpComparison} shows the effective masses of the $T_1^{-+}$ obtained in different ways for $N_\tau = 40$, where the numerical cancellations of the midpoint subtraction are not a problem. The hollow points correspond to using each operator of the $\{ \Gamma, \gamma_4 \} \neq 0$ group individually and the black dots are the effective mass from a GEVP combining these operators together. The different operators are labelled following the convention of \cite{Dudek-2010}. As expected, the GEVP converges faster than any of the eight single operators towards a plateau. The $\times$-shaped points are the effective masses calculated from the single operator satisfying $\{ \Gamma, \gamma_4  \} = 0$. Clearly the latter has a significantly faster convergence to a plateau than any of the results using the other eight operators, thanks to a much better overlap with the ground state. This operator includes the chromo-magnetic component $\mathbb
B_{i} = \epsilon_{ijk} \nabla_j \nabla_k$ via gauge-covariant derivatives $\nabla_i$ and it therefore vanishes in the absence of a gluon background. The explicit gluonic dependence makes it well-suited for sampling a hybrid state. It is clear that, for resolving the ground state, we are not losing any significant information by using the single operator with all the desired symmetries. This is particularly useful when going to very small $N_\tau$ (high temperature), where the midpoint subtraction is much more problematic. %\tcr{This explicit dependence contrasts the other eight operators which attempt to capture gluonic dependence through various spatial derivatives}
\par
We use this single ``best'' operator to extract the ground state effective masses across all temperatures shown in Table \ref{table:Gen2} and the results are presented in \Fig{fig:A1mpProfiles}. At zero temperature~\cite{Liu-2012} the ground state $T_1^{-+}$ has a mass $\approx 0.75$ in lattice units, which is consistent with the results at $T=141$ MeV (the lowest temperature considered here). As the temperature increases, we expect a change in the spectral density function and therefore the effective mass extraction based on the zero-temperature one fails to determine a reliable plateau. This becomes particularly clear going above $T = 176$ MeV, which is the highest temperature we have in this ensemble below $T_c \approx 181$ MeV.
%\begin{figure}
%    \centering
%    \includegraphics[width=0.7\linewidth]{Figs/T1-+_AllTemps.pdf}
%    \caption{Effective masses of the ground state $T_1^{-+}$ $\left( 1^{-+} \right)$ charmonium across all temperatures included in Table. \ref{table:Gen2}.%\tcr{Ryan: Similar comments on marker \& axis label sizes}
%    }
%    \label{fig:T1mpAllTemps}
%\end{figure}
\par
While these results allow us to follow the hybrid charmonium across different temperatures with robust spectroscopy techniques, the time-reversal and $\gamma_4$-anti-commutation restrictions heavily reduce our operator basis. This impedes using a GEVP based on different choices of $\Gamma$ for the $T_1^{-+}$ case and for the channels where there are more than one operator available the resulting correlation matrices are not as complete as in the zero-temperature case. We tackle this issue by using the optimal distillation profiles approach presented in \cite{Urrea-2022}. 
\subsection{Distillation Profiles}
\begin{figure}
    \centering
    \includegraphics[width=0.495\linewidth]{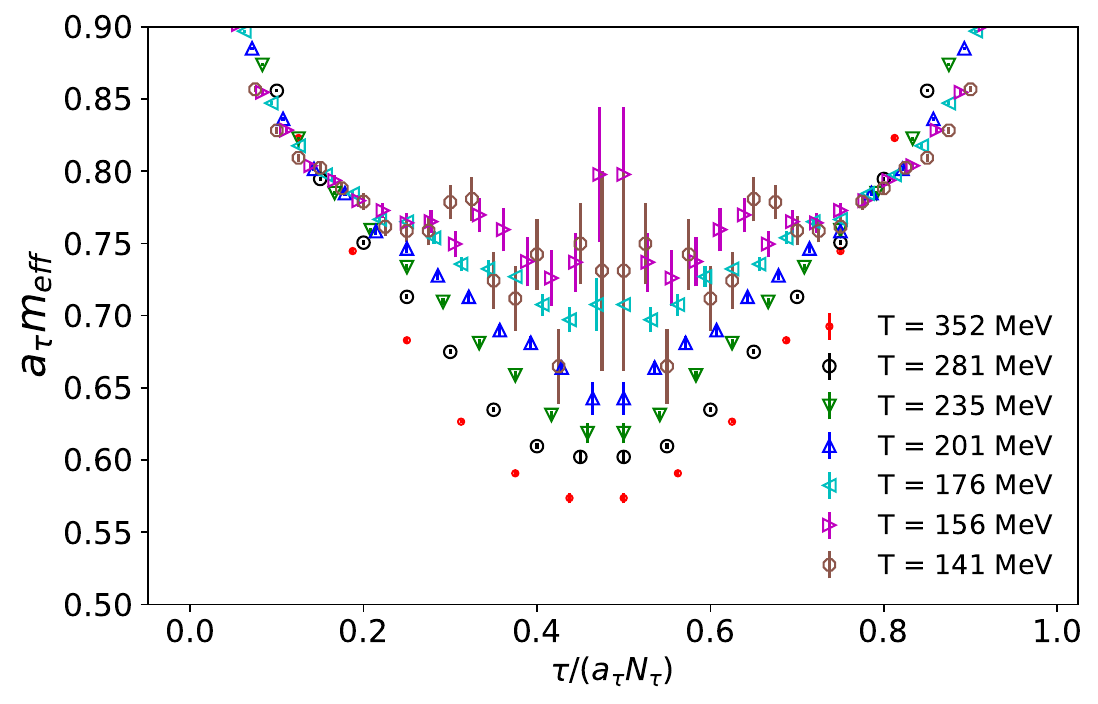}
    \hfill
    \centering
    \includegraphics[width=0.495\linewidth]{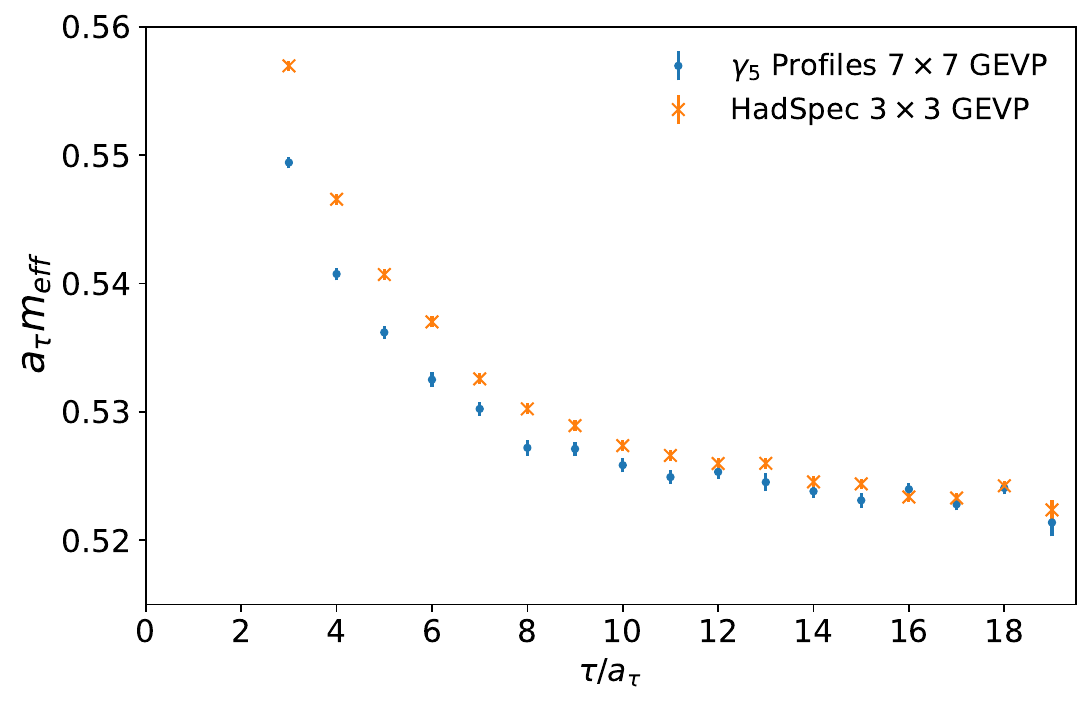}
    \caption{\textbf{Left:} Effective masses of the ground state $T_1^{-+}$ $\left( 1^{-+} \right)$ charmonium across all temperatures included in Table \ref{table:Gen2}. \textbf{Right:} Effective masses of the ground state $A_1^{-+}$ $\left( 0^{-+}  \right)$ charmonium at $N_t = 40$ from two GEVPs: using 3 operators from the available basis and using $\gamma_5$ with 7 different distillation profiles as described in \cite{Urrea-2022}.}
    \label{fig:A1mpProfiles}
\end{figure}

Optimal distillation profiles have been shown to provide a significant and computationally cheap improvement over the standard distillation technique because the vectors involved are used in an optimal way for each state of interest instead of defining an orthogonal projection in every case \cite{Urrea-2022}. They have been used in recent zero-temperature meson spectroscopy calculations \cite{Urrea-2025} and here we apply them for the first time to finite-temperature calculations. Since their inclusion does not change the time-reversal or $\gamma_4$-anti-commutation symmetries of the operator, they are a very simple yet powerful way of increasing the operator basis. We test this first in the $A_1^{-+}$ channel, where we have three operators with all the desired symmetries and can therefore solve a $3\times 3$ GEVP to use as a reference for any improvement the profiles can bring. \Fig{fig:A1mpProfiles} (right pane) shows the ground state effective masses for the $A_1^{-+}$ from two different GEVPs at $N_\tau = 40$; the reference $3\times 3$ one and a $7\times 7$ using $\Gamma = \gamma_5$ with seven different profiles. The $\Gamma = \gamma_5$ operator is also included in the $3\times 3$ GEVP. For the $A_1^{-+}$ ground state, it is clear the use of profiles reduces excited-state contamination more than the inclusion of derivative-based operators together with $\Gamma = \gamma_5$ does. To further reduce excited-state contamination as well as resolve more excitations we can combine both approaches: use all available operators and combine them with distillation profiles, which in this case yields a $21\times 21$ correlation matrix. As explained in \cite{Urrea-2022}, the inclusion of profiles for a fixed $\Gamma$ is significantly cheaper than the inclusion of multiple choices of derivative-based $\Gamma$ since in distillation this requires the calculation of additional elementals to contract with the perambulators. The inclusion of profiles simply requires multiplying the already available elementals from left and right by diagonal matrices containing the distillation profiles. When multiple elementals are already available, the inclusion of profiles comes at a negligible computational cost. This work is underway for the $T_1^{-+}$.

\subsection{Reconstructed Correlators}
\begin{figure}
    \centering
    \overlayText{0.473\linewidth}{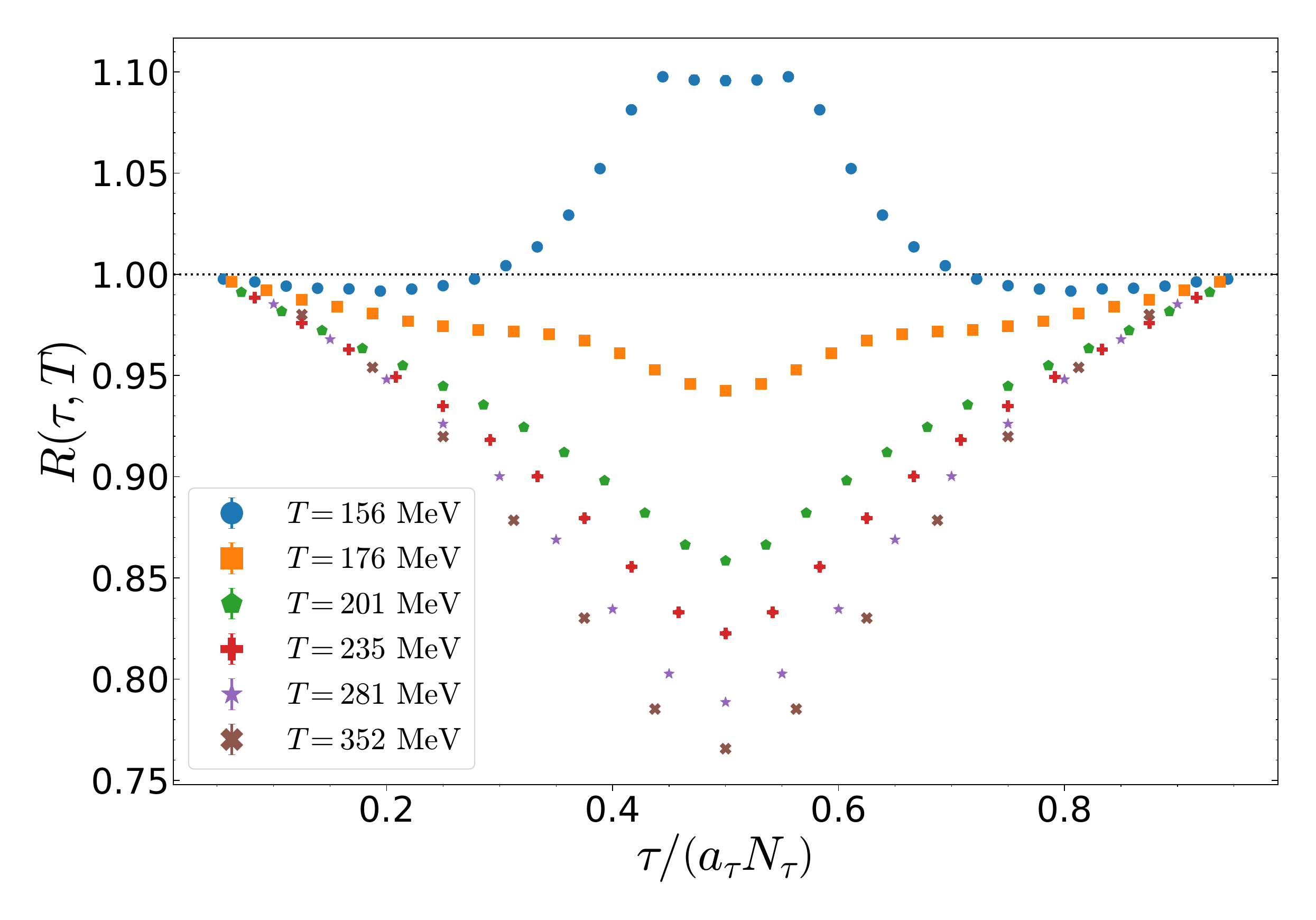}{2.6}{1.8}{$A_1^{-+}$}
    \hfill
    \overlayText{0.473\linewidth}{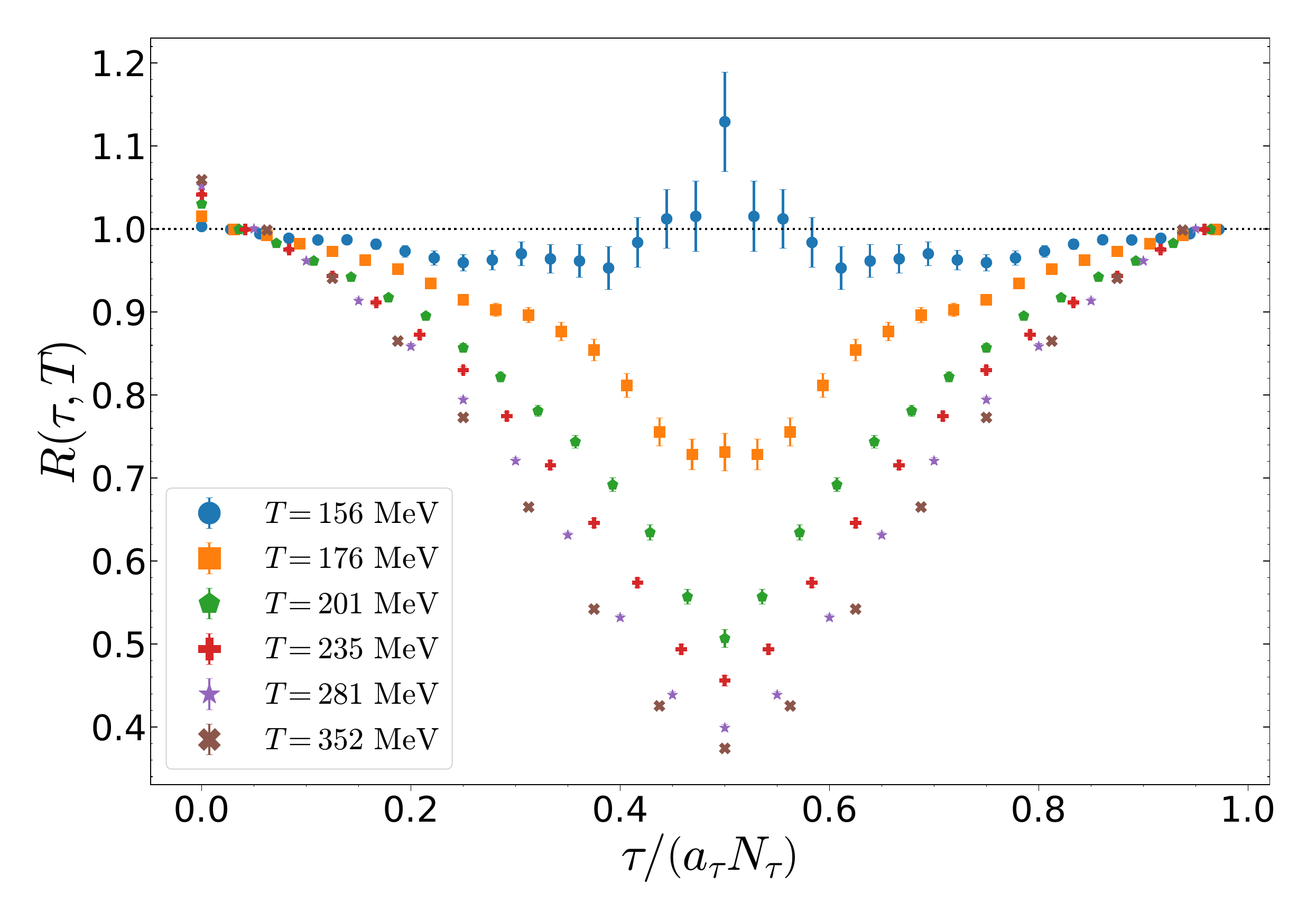}{2.6}{1.8}{$T_1^{-+}$}
    \caption{Ratio of reconstructed correlator (using $N_t=40$ as the zero temperature) to lattice correlator described in \eqnr{eqn:recon:ratio} for the $A_1^{-+}$ ground state (left) and the $T_1^{-+}$ ground state (right). Note that the temperature dependence as indicated by the deviation from unity is much greater for the $T_1^{-+}$ than for the $A_1^{-+}$.}
    \label{fig:recon:A1T1}
\end{figure}
The reconstructed correlator method~\cite{Ding:2012sp,Kelly:2018hsi,Aarts:2022krz}, can give insight into the change in the spectral content of the state without explicit determination of its spectral function. This is done by considering the Euclidean correlator in the following form
\begin{align*}
G(\tau; T) = \int\,d\omega\,K(\tau,\omega; T)\,\rho(\omega;T),
\end{align*}
where $\rho(\omega;T)$ is the spectral function at a given temperature T and the known kernel function is (for bosonic states)
\begin{align*}
K(\tau,\omega;T) = \frac{\cosh{(\omega(\tau - N_\tau/2)})}{\sinh{(\omega\,N_\tau/2})
} = \sum_{n=0}^{m-1}\,\frac{\cosh{\left[\omega(\tau+n\,N_\tau - m\,N_\tau /2)\right]}}{\sinh{(\omega\,m\,N_\tau/2)}}.
\end{align*}
Here $\tau,\,n,\,m$ are positive integers and $N_\tau = 1/(a_\tau\,T)$. Hence we may relate the kernels at two temperatures $T_0$ and $T=m\,T_0$ by
\begin{align*}
K(\tau,\omega;T) = \sum_{n=0}^{m-1}\,K(\tau + a_\tau\,n\,N_\tau, \omega; T_0).
\end{align*}
We can then construct the reconstructed correlator at a higher temperature using a summation of the zero-temperature correlator
\begin{align*}
    G_R(\tau;T,T_0) = \sum_{n=0}^{m-1}\,G(\tau + n / T; T_0).
\end{align*}
This requires that $T/T_0$ is an integer; as this is not usually the case the zero-temperature correlator is \enquote{padded} with extra data points until an integer ratio is obtained. For truly zero-temperature heavy meson correlators this has negligible effect if padding is done with zeroes or the minimum value of the correlator.
\par
The reconstruction allows an examination of the \textit{change} to the spectral function $\rho(\omega,T)$ without requiring a spectral reconstruction but offers no insight into what the changes are. To examine the change we consider a ratio of the reconstructed to lattice correlators
\begin{align}
R(\tau, T) = \frac{G_{R}(\tau, T; T_0)}{G(\tau, T)},
\label{eqn:recon:ratio}
\end{align}
such that small shifts from one indicate only small changes in the spectral function.
\par
We present this ratio in \Fig{fig:recon:A1T1} where the $A_1^{-+}$ ground state (left) is compared to the $T_1^{-+}$ ground state (right). It is immediately clear that the $T_1^{-+}$ ground state is much more affected by the temperature than the $A_1^{-+}$ ground state as the correlator ratio is significantly further from one at all temperatures above $T\approx176$ MeV. The rising upwards at the centre of the $T=156$ MeV ratio for the $A_1^{-+}$ is attributed to the use of the $N_\tau=40$ ensemble as the zero-temperature correlator in the ratio rather than the actual zero-temperature ($N_t=128$) ensemble. This has a two-fold effect: firstly the $N_\tau=40$ $A_1^{-+}$ correlator is not very exponentially suppressed in the middle of the lattice yet (indeed, it is two orders of magnitude larger than the corresponding $T_1^{-+}$ correlator) and secondly the $N_\tau=36$ requires the most \enquote{padding} to form an integer ratio. Both of these problems may be resolved through use of the $N_\tau=128$ ensemble and work in this direction is in progress.

\section{Conclusions and Outlook}
In this work we successfully applied, for the first time, the distillation technique together with a HadSpec's wide basis of meson operators for spectroscopy at finite temperature. While most of the process involved in this approach remains the same as in the zero-temperature case, e.g.~calculation of perambulators and elementals, the finite-temperature setting includes two additional symmetries which must be carefully dealt with at the meson operator level before effective masses can be reliably extracted from a GEVP; time-reversal and $\gamma_4$-anti-commutation. While the most straightforward way of doing this is further grouping the available operators based on these two symmetries, this effectively reduces the number of operators available for each GEVP. For the $T_1^{-+}$ this results in just one operator with all the desired symmetries, impeding the use of a GEVP involving more than one operator. Other channels such as the $A_1^{-+}$ suffer a reduction from $12$ to $3$. We tackled this issue by using the recently proposed approach of optimal distillation profiles. The introduction of the profiles does not break either of the two additional symmetries and therefore allows an increase the basis size. For the case of $A_1^{-+}$ we observed that one choice of $\Gamma$ with different profiles resulted in ground state effective masses with less excited state contamination than the use of multiple $\Gamma$. We plan on combining both approaches to fully exploit a more complete basis of operators to get physically useful information at earlier time separations of the correlation functions, which is vital when going to high enough temperatures where the temporal extent is very short. Nonetheless, with the standard approach used here we could map the effective masses of the hybrid charmonium across different temperatures. As a next step, we plan to repeat and extend these calculations on \textsc{Fastsum}'s Generation 3 ensembles~\cite{Skullerud:2025xva}. Their physical characteristics are very similar to Generation 2 however they have twice as many temporal points at each temperature.
\par
The reconstructed correlator ratio reveals that the $T_1^{-+}$ ground state is more strongly affected by temperature than the corresponding $A_1^{-+}$ ground state. This is an important check as it distinguishes temperature effects from known temporal length effects which are present in e.g.~the effective mass. The use of $N_\tau=40$ as the zero-temperature ensemble is seen to not be a good substitute for zero temperature and this will be corrected using the $N_\tau = 128$ ensemble. It would be interesting to apply a spectral function reconstruction method such as the Maximum Entropy~\cite{Aarts:2010ek,Aarts:2011sm} or Bayesian reconstruction~\cite{Burnier:2013nla,Rothkopf:2022ctl} methods to directly investigate the spectral content of the $T_1^{-+}$ at finite temperature.
\section*{Software \& Data}
The error analysis is done using the $\Gamma$-method~\cite{Wolff-2004, Wolff-2007, Schaefer-2011} with automatic differentiation~\cite{Ramos-2019} via the \textbf{pyerrors} library~\cite{Joswig-2023}. We thank the Hadron Spectrum Collaboration for the use of its code framework, particularly the \textbf{Chroma}~\cite{Edwards:2004sx} software suite and the \textbf{Redstar}~\cite{10.1145/3592979.3593409} package. The gauge field ensembles are publicly available~\cite{aarts_2023_8403827} and we anticipate making other data available after a future publication.

\section*{Acknowledgements}
This work used the computing resources of the Irish Centre
for High-End Computing (ICHEC) and was supported by RIT (Research IT, Trinity College Dublin). R.B., R.C., J.A.U.-N.\ and S.R.\ acknowledge support from a Research Ireland (Science Foundation Ireland) Frontiers for the Future Project award with grant number SFI-21/FFP-P/10186.
\bibliographystyle{JHEP_arXiv}
\bibliography{refs}

@article{Liu-2012,
  title = {Excited and exotic charmonium spectroscopy from lattice QCD},
  volume = {2012},
  ISSN = {1029-8479},
  url = {http://dx.doi.org/10.1007/JHEP07(2012)126},
  DOI = {10.1007/jhep07(2012)126},
  number = {7},
  journal = {Journal of High Energy Physics},
  publisher = {Springer Science and Business Media LLC},
  author = {Liu,  Liuming and Moir,  Graham and Peardon,  Michael and Ryan,  Sinéad M. and Thomas,  Christopher E. and Vilaseca,  Pol and Dudek,  Jozef J. and Edwards,  Robert G. and Joó,  Bálint and Richards,  David G.},
  year = {2012},
  month = jul 
}

@article{Dudek-2008,
  title = {Charmonium excited state spectrum in lattice QCD},
  author = {Dudek, Jozef J. and Edwards, Robert G. and Mathur, Nilmani and Richards, David G.},
  journal = {Phys. Rev. D},
  volume = {77},
  issue = {3},
  pages = {034501},
  numpages = {22},
  year = {2008},
  month = {Feb},
  publisher = {American Physical Society},
  doi = {10.1103/PhysRevD.77.034501},
  url = {https://link.aps.org/doi/10.1103/PhysRevD.77.034501}
}

@article{Peardon-2009,
  title = {Novel quark-field creation operator construction for hadronic physics in lattice QCD},
  volume = {80},
  ISSN = {1550-2368},
  url = {http://dx.doi.org/10.1103/PhysRevD.80.054506},
  DOI = {10.1103/physrevd.80.054506},
  number = {5},
  journal = {Physical Review D},
  publisher = {American Physical Society (APS)},
  author = {Peardon,  Michael and Bulava,  John and Foley,  Justin and Morningstar,  Colin and Dudek,  Jozef and Edwards,  Robert G. and Joó,  Bálint and Lin,  Huey-Wen and Richards,  David G. and Juge,  Keisuke Jimmy},
  year = {2009},
  month = sep 
}

@article{Luscher-1990,
  title = {How to calculate the elastic scattering matrix in two-dimensional quantum field theories by numerical simulation},
  volume = {339},
  ISSN = {0550-3213},
  url = {http://dx.doi.org/10.1016/0550-3213(90)90540-T},
  DOI = {10.1016/0550-3213(90)90540-t},
  number = {1},
  journal = {Nuclear Physics B},
  publisher = {Elsevier BV},
  author = {L\"{u}scher,  Martin and Wolff,  Ulli},
  year = {1990},
  month = jul,
  pages = {222–252}
}

@article{Blossier-2009,
  title = {On the generalized eigenvalue method for energies and matrix elements in lattice field theory},
  volume = {2009},
  ISSN = {1029-8479},
  url = {http://dx.doi.org/10.1088/1126-6708/2009/04/094},
  DOI = {10.1088/1126-6708/2009/04/094},
  number = {04},
  journal = {Journal of High Energy Physics},
  publisher = {Springer Science and Business Media LLC},
  collaboration = {ALPHA},
  author = {Blossier,  Benoit and Morte,  Michele Della and Hippel,  Georg von and Mendes,  Tereza and Sommer,  Rainer},
  year = {2009},
  month = apr,
  pages = {094–094}
}

@article{Bailas-2018,
  title = {Some hadronic parameters of charmonia in $\text{N}_{\text {f}}=2$ lattice {QCD}},
  volume = {78},
  ISSN = {1434-6052},
  url = {http://dx.doi.org/10.1140/epjc/s10052-018-6495-4},
  DOI = {10.1140/epjc/s10052-018-6495-4},
  number = {12},
  journal = {The European Physical Journal C},
  publisher = {Springer Science and Business Media LLC},
  author = {Bailas,  Gabriela and Blossier,  Benoît and Morénas,  Vincent},
  year = {2018},
  month = {December} 
}

@article{Umeda-2007,
  title = {Constant contribution in meson correlators at finite temperature},
  author = {Umeda, Takashi},
  journal = {Phys. Rev. D},
  volume = {75},
  issue = {9},
  pages = {094502},
  numpages = {11},
  year = {2007},
  month = {May},
  publisher = {American Physical Society},
  doi = {10.1103/PhysRevD.75.094502},
  url = {https://link.aps.org/doi/10.1103/PhysRevD.75.094502}
}

@article{Ohno-2011,
  title = {Charmonium spectral functions with the variational method in zero and finite temperature lattice QCD},
  author = {Ohno, H. and Aoki, S. and Ejiri, S. and Kanaya, K. and Maezawa, Y. and Saito, H. and Umeda, T.},
  collaboration = {WHOT-QCD},
  journal = {Phys. Rev. D},
  volume = {84},
  issue = {9},
  pages = {094504},
  numpages = {13},
  year = {2011},
  month = {Nov},
  publisher = {American Physical Society},
  doi = {10.1103/PhysRevD.84.094504},
  url = {https://link.aps.org/doi/10.1103/PhysRevD.84.094504}
}

@article{Arikawa-2025,
  title = {Glueball mass spectrum at finite temperature revisited: Constant contribution in glueball correlators in the deconfinement phase},
  author = {Arikawa, Toshizo and Sakai, Keita and Sasaki, Shoichi},
  journal = {Phys. Rev. D},
  volume = {112},
  issue = {1},
  pages = {014506},
  numpages = {14},
  year = {2025},
  month = {Jul},
  publisher = {American Physical Society},
  doi = {10.1103/3qy4-b8ym},
  url = {https://link.aps.org/doi/10.1103/3qy4-b8ym}
}

@article{Urrea-2025,
  title = {S-wave flavor-singlet meson mixing in QCD with light and charm quarks},
  author = {Urrea-Ni\~no, Juan Andr\'es and H\"ollwieser, Roman and Knechtli, Francesco and Korzec, Tomasz and Finkenrath, Jacob and Peardon, Michael},
  journal = {Phys. Rev. D},
  volume = {112},
  issue = {7},
  pages = {074502},
  numpages = {14},
  year = {2025},
  month = {Oct},
  publisher = {American Physical Society},
  doi = {10.1103/tcbk-x4f1},
  url = {https://link.aps.org/doi/10.1103/tcbk-x4f1}
}

@article{Dudek-2010,
  title = {Toward the excited meson spectrum of dynamical QCD},
  author = {Dudek, Jozef J. and Edwards, Robert G. and Peardon, Michael J. and Richards, David G. and Thomas, Christopher E.},
  collaboration = {Hadron Spectrum},
  journal = {Phys. Rev. D},
  volume = {82},
  issue = {3},
  pages = {034508},
  numpages = {24},
  year = {2010},
  month = {Aug},
  publisher = {American Physical Society},
  doi = {10.1103/PhysRevD.82.034508},
  url = {https://link.aps.org/doi/10.1103/PhysRevD.82.034508}
}

@article{Urrea-2022,
  title = {Optimizing creation operators for charmonium spectroscopy on the lattice},
  author = {Knechtli, Francesco and Korzec, Tomasz and Peardon, Michael and Urrea-Ni\~no, Juan Andr\'es},
  journal = {Phys. Rev. D},
  volume = {106},
  issue = {3},
  pages = {034501},
  numpages = {19},
  year = {2022},
  month = {Aug},
  publisher = {American Physical Society},
  doi = {10.1103/PhysRevD.106.034501},
  url = {https://link.aps.org/doi/10.1103/PhysRevD.106.034501}
}

@article{Aarts:2014nba,
    author = "Aarts, Gert and Allton, Chris and Amato, Alessandro and Giudice, Pietro and Hands, Simon and Skullerud, Jon-Ivar",
    title = "{Electrical conductivity and charge diffusion in thermal QCD from the lattice}",
    eprint = "1412.6411",
    archivePrefix = "arXiv",
    primaryClass = "hep-lat",
    reportNumber = "HIP-2014-34-TH, INT-PUB-14-060, MS-TP-14-40",
    doi = "10.1007/JHEP02(2015)186",
    journal = "JHEP",
    volume = "02",
    pages = "186",
    year = "2015"
}

@article{Aarts:2020vyb,
    author = "G. Aarts and C. Allton and J. Glesaaen and S. Hands and B. Jäger and S. Kim and M. P. Lombardo and A. A. Nikolaev and S. M. Ryan and J. -I. Skullerud and L. -K. Wu",
    title = "{Properties of the QCD thermal transition with Nf=2+1 flavors of Wilson quark}",
    eprint = "2007.04188",
    archivePrefix = "arXiv",
    primaryClass = "hep-lat",
    doi = "10.1103/PhysRevD.105.034504",
    journal = "Phys. Rev. D",
    volume = "105",
    number = "3",
    pages = "034504",
    year = "2022"
}

@article{Wolff-2004,
title = {Monte Carlo errors with less errors},
journal = {Computer Physics Communications},
volume = {156},
number = {2},
pages = {143-153},
year = {2004},
issn = {0010-4655},
doi = {https://doi.org/10.1016/S0010-4655(03)00467-3},
url = {https://www.sciencedirect.com/science/article/pii/S0010465503004673},
author = {Ulli Wolff}
}

@article{Wolff-2007,
title = {Erratum to “Monte Carlo errors with less errors” [Comput. Phys. Comm. 156 (2004) 143–153]},
journal = {Computer Physics Communications},
volume = {176},
number = {5},
pages = {383},
year = {2007},
issn = {0010-4655},
doi = {https://doi.org/10.1016/j.cpc.2006.12.001},
url = {https://www.sciencedirect.com/science/article/pii/S0010465506004322},
author = {Ulli Wolff}
}

@article{Schaefer-2011,
title = {Critical slowing down and error analysis in lattice QCD simulations},
journal = {Nuclear Physics B},
volume = {845},
number = {1},
pages = {93-119},
year = {2011},
issn = {0550-3213},
doi = {https://doi.org/10.1016/j.nuclphysb.2010.11.020},
url = {https://www.sciencedirect.com/science/article/pii/S0550321310006188},
author = {Stefan Schaefer and Rainer Sommer and Francesco Virotta}
}

@article{Ramos-2019,
title = {Automatic differentiation for error analysis of Monte Carlo data},
journal = {Computer Physics Communications},
volume = {238},
pages = {19-35},
year = {2019},
issn = {0010-4655},
doi = {https://doi.org/10.1016/j.cpc.2018.12.020},
url = {https://www.sciencedirect.com/science/article/pii/S0010465519300013},
author = {Alberto Ramos}
}

@article{Joswig-2023,
title = {pyerrors: A python framework for error analysis of Monte Carlo data},
journal = {Computer Physics Communications},
volume = {288},
pages = {108750},
year = {2023},
eprint = "2209.14371",
archivePrefix = "arXiv",
primaryClass = "hep-lat",
issn = {0010-4655},
doi = {https://doi.org/10.1016/j.cpc.2023.108750},
url = {https://www.sciencedirect.com/science/article/pii/S0010465523000954},
author = {Fabian Joswig and Simon Kuberski and Justus T. Kuhlmann and Jan Neuendorf}
}

@article{Brambilla-2020,
title = {The XYZ states: Experimental and theoretical status and perspectives},
journal = {Physics Reports},
volume = {873},
pages = {1-154},
year = {2020},
note = {The XYZ states: experimental and theoretical status and perspectives},
issn = {0370-1573},
doi = {https://doi.org/10.1016/j.physrep.2020.05.001},
url = {https://www.sciencedirect.com/science/article/pii/S0370157320301915},
author = {Nora Brambilla and Simon Eidelman and Christoph Hanhart and Alexey Nefediev and Cheng-Ping Shen and Christopher E. Thomas and Antonio Vairo and Chang-Zheng Yuan}
}

@article{Godfrey-2008,
   author = "Godfrey, Stephen and Olsen, Stephen L.",
   title = "The Exotic XYZ Charmonium-Like Mesons", 
   journal= "Annual Review of Nuclear and Particle Science",
   year = "2008",
   volume = "58",
   number = "Volume 58, 2008",
   pages = "51-73",
   doi = "https://doi.org/10.1146/annurev.nucl.58.110707.171145",
   url = "https://www.annualreviews.org/content/journals/10.1146/annurev.nucl.58.110707.171145",
   publisher = "Annual Reviews",
   issn = "1545-4134",
   type = "Journal Article",
  }

@article{Shi-2024,
  title = {Decays of ${1}^{\ensuremath{-}+}$ charmoniumlike hybrid using lattice QCD},
  author = {Shi, Chunjiang and Chen, Ying and Gong, Ming and Jiang, Xiangyu and Liu, Zhaofeng and Sun, Wei},
  journal = {Phys. Rev. D},
  volume = {109},
  issue = {9},
  pages = {094513},
  numpages = {17},
  year = {2024},
  month = {May},
  publisher = {American Physical Society},
  doi = {10.1103/PhysRevD.109.094513},
  url = {https://link.aps.org/doi/10.1103/PhysRevD.109.094513}
}

@misc{aarts_2023_8403827,
  author       = {Aarts, Gert and
                  Allton, Christopher and
                  Amato, Alessandro and
                  Bignell, Ryan and
                  Burns, Timothy J. and
                  De Boni, Davide and
                  Evans, Wynne and
                  Glesaaen, Aleksandra and
                  Giudice, Pietro and
                  Hands, Simon and
                  Harris, Tim and
                  Jaeger, Benjamin and
                  Kelly, Aoife and
                  Kim, Seyong and
                  Lombardo, Maria Paola and
                  Nikolaev, Aleksandr and
                  Quinn, Ryan and
                  Rothkopf, Alexander and
                  Ryan, Sinead and
                  Skullerud, Jon-Ivar and
                  Spriggs, Thomas and
                  Wu, Liang-Kai},
  title        = {FASTSUM Generation 2 Anisotropic Thermal Lattice
                   QCD Gauge Ensembles
                  },
  month        = oct,
  year         = 2023,
  publisher    = {Zenodo},
  doi          = {\href{https://doi.org/10.5281/zenodo.8403827}{ 10.5281/zenodo.8403827}},
  url          = {https://doi.org/10.5281/zenodo.8403827},
}

@inproceedings{10.1145/3592979.3593409,
author = {Chen, Jie and Edwards, Robert G. and Mao, Weizhen},
title = {Graph Contractions for Calculating Correlation Functions in Lattice QCD},
year = {2023},
isbn = {9798400701900},
publisher = {Association for Computing Machinery},
address = {New York, NY, USA},
url = {https://doi.org/10.1145/3592979.3593409},
doi = {10.1145/3592979.3593409},
booktitle = {Proceedings of the Platform for Advanced Scientific Computing Conference},
articleno = {10},
numpages = {10},
location = {Davos, Switzerland},
series = {PASC '23}
}

@article{Edwards:2004sx,
    author = "Edwards, Robert G. and Joo, Balint",
    editor = "Bodwin, Geoffrey T. and Sinclair, D. K. and Eichten, E. and Holmgren, D. and Kronfeld, Andreas S. and Mackenzie, P. and Okamoto, M. and Simone, J. and El-Khadra, Aida X.",
    collaboration = "SciDAC, LHPC, UKQCD",
    title = "{The Chroma software system for lattice QCD}",
    eprint = "hep-lat/0409003",
    archivePrefix = "arXiv",
    reportNumber = "JLAB-THY-04-54",
    doi = "10.1016/j.nuclphysbps.2004.11.254",
    journal = "Nucl. Phys. B Proc. Suppl.",
    volume = "140",
    pages = "832",
    year = "2005"
}

@article{Skullerud:2025xva,
    author={Jon-Ivar Skullerud and Gert Aarts and Chris Allton and M. Naeem Anwar and Ryan Bignell and Tim Burns and Simon Hands and Rachel Horohan D'Arcy and Ben Jäger and Seyong Kim and Alan Kirby and Maria Paola Lombardo and Seung-Il Nam and Sinéad M. Ryan and Antonio Smecca},
    title = "{Approaching the continuum with anisotropic lattice thermodynamics}",
    eprint = "2510.02954",
    archivePrefix = "arXiv",
    primaryClass = "hep-lat",
    doi = "10.1016/j.jspc.2025.100258",
    journal = "J. Subatomic Part. Cosmol.",
    volume = "4",
    pages = "100258",
    year = "2025"
}

@article{Kelly:2018hsi,
    author = "Kelly, Aoife and Rothkopf, Alexander and Skullerud, Jon-Ivar",
    title = "{Bayesian study of relativistic open and hidden charm in anisotropic lattice QCD}",
    eprint = "1802.00667",
    archivePrefix = "arXiv",
    primaryClass = "hep-lat",
    doi = "10.1103/PhysRevD.97.114509",
    journal = "Phys. Rev. D",
    volume = "97",
    number = "11",
    pages = "114509",
    year = "2018"
}

@article{Ding:2012sp,
    author = "Ding, H. T. and Francis, A. and Kaczmarek, O. and Karsch, F. and Satz, H. and Soeldner, W.",
    title = "{Charmonium properties in hot quenched lattice QCD}",
    eprint = "1204.4945",
    archivePrefix = "arXiv",
    primaryClass = "hep-lat",
    reportNumber = "BI-TP-2012-13",
    doi = "10.1103/PhysRevD.86.014509",
    journal = "Phys. Rev. D",
    volume = "86",
    pages = "014509",
    year = "2012"
}

@article{Aarts:2022krz,
    author = {Aarts, Gert and Allton, Chris and Bignell, Ryan and Burns, Timothy J. and Garc\'\i{}a-Mascaraque, Sergio Chaves and Hands, Simon and J\"ager, Benjamin and Kim, Seyong and Ryan, Sin\'ead M. and Skullerud, Jon-Ivar},
    title = "{Open charm mesons at nonzero temperature: results in the hadronic phase from lattice QCD}",
    eprint = "2209.14681",
    archivePrefix = "arXiv",
    primaryClass = "hep-lat",
    month = "9",
    year = "2022"
}

@article{Burnier:2013nla,
    author = "Burnier, Yannis and Rothkopf, Alexander",
    title = "Bayesian Approach to Spectral Function Reconstruction for Euclidean Quantum Field Theories",
    eprint = "1307.6106",
    archivePrefix = "arXiv",
    primaryClass = "hep-lat",
    doi = "10.1103/PhysRevLett.111.182003",
    journal = "Phys. Rev. Lett.",
    volume = "111",
    pages = "182003",
    year = "2013"
}

@article{Rothkopf:2022ctl,
    author = "Rothkopf, Alexander",
    title = "{Bayesian inference of real-time dynamics from lattice QCD}",
    eprint = "2208.13590",
    archivePrefix = "arXiv",
    primaryClass = "hep-lat",
    doi = "10.3389/fphy.2022.1028995",
    journal = "Front. Phys.",
    volume = "10",
    pages = "1028995",
    year = "2022"
}

@article{Aarts:2010ek,
    author = "Aarts, G. and Kim, S. and Lombardo, M. P. and Oktay, M. B. and Ryan, S. M. and Sinclair, D. K. and Skullerud, J. -I.",
    title = "{Bottomonium above deconfinement in lattice nonrelativistic QCD}",
    eprint = "1010.3725",
    archivePrefix = "arXiv",
    primaryClass = "hep-lat",
    reportNumber = "ANL-HEP-PR-10-55",
    doi = "10.1103/PhysRevLett.106.061602",
    journal = "Phys. Rev. Lett.",
    volume = "106",
    pages = "061602",
    year = "2011"
}

@article{Aarts:2011sm,
    author = "Aarts, G. and Allton, C. and Kim, S. and Lombardo, M. P. and Oktay, M. B. and Ryan, S. M. and Sinclair, D. K. and Skullerud, J. I.",
    title = "{What happens to the $\Upsilon$ and $\eta_b$ in the quark-gluon plasma? Bottomonium spectral functions from lattice QCD}",
    eprint = "1109.4496",
    archivePrefix = "arXiv",
    primaryClass = "hep-lat",
    doi = "10.1007/JHEP11(2011)103",
    journal = "JHEP",
    volume = "11",
    pages = "103",
    year = "2011"
}

@article{Peardon-2004,
  title = {Analytic smearing of $\mathrm{SU}(3)$ link variables in lattice QCD},
  author = {Morningstar, Colin and Peardon, Mike},
  journal = {Phys. Rev. D},
  volume = {69},
  issue = {5},
  pages = {054501},
  numpages = {9},
  year = {2004},
  month = {Mar},
  publisher = {American Physical Society},
  doi = {10.1103/PhysRevD.69.054501},
  url = {https://link.aps.org/doi/10.1103/PhysRevD.69.054501}
}

\end{document}